\documentclass[12pt]{article}
\begin{document}
\title{ {\bf  A generating functional for ultrasoft amplitudes in hot QCD} }
\author{F. Guerin \\ 
  Institut Non Lineaire de Nice,  1361 route des Lucioles, 06560 Valbonne, France } \date{}
\maketitle
\begin{abstract}
The effective amplitudes for gluon momenta $p \ll   gT$ in hot QCD exhibit damping as a result of collisions. The whole set of $n$-point amplitudes is shown to be generated from one functional ${\mathcal K}_{\mu \nu}(X,Y; A)$, in addition to the induced current $j_{\mu}(X; A)$.
\\ 

\end{abstract}

The ultrasoft amplitudes are the effective amplitudes that are obtained in pure gauge theory at high temperature, when one integrates out the scales  $(T)^{-1}$ and  $(gT)^{-1}$  \cite{Bod,BI1,AY}. The colour collective excitations,  which describe the physics at such a large wavelength, may be  created by a weak external disturbance. The ultrasoft fields obey the Maxwell equations \cite{BI1}
 \begin{equation}
 D^{\nu} F_{\mu\nu}(X) = j_{\mu}^{ext}(X) + j_{\mu}^{ind}(X)
 \end{equation}
where the induced current is the response of the plasma to the initial perturbation $j_{\mu}^{ext}(X)$.
 \begin{equation}
 j_{\mu}^{a \ ind}(X) =m_D^2 \ \int_v \ v_{\mu} \ W^a(X,{\bf v})
 \label{1}
 \end{equation}
\begin{equation}
(v.D_X+\hat{C})^{ab}\ W^b(X,{\mathbf v}) = {\mathbf v}.{\mathbf E}^a(X)
\label{2}
\end{equation}
with $v^{\mu}(v_0=1,{\mathbf v}), ({\mathbf v}^2=1)$ and $\int_v=\int {d \Omega_v /{4\pi}}$ ;  $ m_D^2= g^2NT^2/ 3$ and $\hat{C}$ is a collision term  \cite{Bod,BI1,AY}
\begin{equation}
\hat{C} \  W^b(X, \mathbf{v})  =\int_{v'}  \ C({\mathbf{v}, \mathbf{v'}}) W^b(X,{\mathbf{v'}})
\end{equation}
$\hat{C}$ is a symmetric operator in $\mathbf{v}$ space,  local in $X$ and blind to colour, with positive eigenvalues except for a zero mode
\begin{equation}
\int _{v'} C({\mathbf{v}, \mathbf{v'}}) = 0 \  .
\label{3}
\end{equation}
The gluons $p\sim T$ take part in the collective motion, the gluons $p\sim g T$ are exchanged in the collision process.

The solution to Eqs.(\ref{1})(\ref{2}) may be written in terms of the retarded Green function
\begin{equation}
i\ (v.D_x + \hat{C}) \ G_{ret}( X, Y;A ;{\mathbf v}, {\mathbf v'})  = 
 \delta^{(4)}(X-Y) \ \delta_{S_2} ({\mathbf v} - {\mathbf v'})
\label{4a} 
\end{equation} 
with $ G_{ret} ( X, Y)  =  0$  for $ X_0 < Y_0$. One writes
\begin{equation}
 {\mathbf v'}.{\mathbf E}(Y)  = \partial_{Y_O}(v'^{\mu} A_{\mu}) - v'.D_Y A_0(Y)
 \end{equation}
and, with Eq.(\ref{3}),  one obtains \cite{BI3,FG3}
 \begin{eqnarray}
 j_{\mu}^{a\ ind}(X; A) =  m_D^2[- g_{0\mu} A_0^a(X) \nonumber \\
 + i  \int_{v,v'}  \int d^4Y \ v_{\mu} \ 
  G_{ret}^{ab}( X, Y ;{\mathbf v}, {\mathbf v'})\  \partial_{Y_0} (v'.A^b(Y)) \ ]
  \end{eqnarray}
The (one-particle-irreducible) $n$-point retarded ultrasoft amplitudes are derived from the induced current \cite{BI1,BI3}
 \begin{equation}
 g^{n-2} G_{ret}(X ; X_1 \cdots X_{n-1}) = 
 {\partial^{n-1} \over{\partial A(X_{n-1}) \cdots
 \partial A(X_1)}} \ \ j_{\mu}^{ind}(X; A) \ \vert_{A=0}
 \label{4}
 \end{equation}
The identity \cite{BI3,FG3}
 \begin{eqnarray}
 \lefteqn{{\partial G_{ret}( X, Y ; A ; {\mathbf v}, {\mathbf v'}) \over{\partial A_c^{\rho}(X_1)}}}  \nonumber \\ & &  =
 \int_{v"} G_{ret}( X, X_1; A ;{\mathbf v}, {\mathbf v"}) \ g T^c
 {v"}_{\rho} \ G_{ret}( X_1, Y; A ;{\mathbf v"}, {\mathbf v'})
 \label{5}
 \end{eqnarray}
shows that the effective ultrasoft amplitudes have a tree-like structure, as the addition of a gluon leg amounts to the insertion of the leg along the propagator (in Eq.(\ref{5}), the time arguments satisfy $X^0 \geq X_1^0 \geq Y^0$ and $T^c$ is in
 the adjoint representation). After differentiation, the Green function that enters the amplitudes is  $G_{ret}(X,Y; A=0;{\mathbf v}, {\mathbf v'})$, i.e. in momentum space
\begin{eqnarray}
 (v.p + i \hat{C})\ G_{ret}(p;{\mathbf v}, {\mathbf v'}) ={\cal I} \nonumber \\ 
  \ \ G_{ret}(p;{\mathbf v}, {\mathbf v'}) = (v.p + i \hat{C})^{-1}_{\ \ v v'}
 \label{6}
 \end{eqnarray}
The amplitudes exhibit damping (i.e. colour relaxation), as a result of collisions. $\hat C$ introduces the scale $g^2T\ln 1/g$ \cite{Bod,BI1,AY}. This damping is compatible with gauge symmetry, because the operator $\hat C$ is local in $X$, blind to color, and has the property (\ref{3}). These amplitudes obey tree-like Ward identities \cite{FG3}.

The retarded amplitudes (\ref{4}) are only a subset of the $n$-point amplitudes, as $X^0$ is the later time. The purpose of this letter is to show that the whole set of $n$-point amplitudes may be generated if one introduces the functional
\begin{eqnarray}
\lefteqn{{\mathcal K}_{\mu \nu}^{a b}(X,Y;A) =m_D^2 \ T}  \nonumber \\ & &
\int_{v,v'}[\  v_{\mu}G_{ret}^{ab}(X,Y; A;{\mathbf v}, {\mathbf v'}) v_{\nu}' \ + \ v_{\nu}G_{ret}^{ba}(Y,X; A;{\mathbf v}, {\mathbf v'}) v_{\mu}' \ ]
\label{7}
\end{eqnarray}
Because of damping, the effective theory has a time arrow. Among the possible real-time formalisms for field theory in thermal equilibrium \cite{MLB,Au1,Chou}, this feature selects the Retarded-Advanced basis ($R/A$), where, because of causality, the propagator keeps its two-component structure $G_{ret}, G_{adv}$ with $G_{ret}(p; {\mathbf v}, {\mathbf v'}) $ as in Eq.(\ref{6}) and $ G_{adv}(p; {\bf v}, {\bf v'})=G_{ret}^{\dagger}(p; {\bf v}, {\bf v'})$. The retarded $n$-point functions are, in fact, common to two bases, the $R/A$ basis and the Keldysh basis, where they correspond to one Keldysh index ``2'' and the other indices  ``1''. There exists a Bogoliubov transformation between the two bases \cite {vWK}.

The ultrasoft amplitudes are such that ${\mathcal K}_{\mu \nu}(X,Y;A)$, defined in Eq.(\ref{7}), is the generating functional for the amplitude with two indices ``2''. Moreover, the amplitudes with more than two indices ``2'' vanish. In the following, it is shown that this prescription generates a consistent set of $n$-point amplitudes in the $R/A$ basis, and these obey the expected Ward identities.

Before addressing the case of the amplitudes with $p \ll gT$, one needs to recall some general properties of the  bases and their connection. In both bases, all momenta are incoming the $n$-point amplitudes, i.e. $\sum_{l=1}^n p_l =0$. The Keldysh basis is a rotated representation of the Closed-Time-Path representation \cite{Chou}, the Keldysh indices are $k_i=1,2$. In the $R/A$ basis \cite{Au1}, an incoming momentum may be of type $R$ or of type $A$, its incoming energy is $p_l^0+i\epsilon_l$,   $\epsilon_l>0$ type $R$, $\epsilon_l<0$ type $A$ with  $\sum_{l=1}^n p_l^0 =0$ and $\sum_{l=1}^n \epsilon_l =0$.

The relations between the (amputated) $n$-point functions in the Keldysh basis and in the $R/A$ basis were obtained in perturbation theory in Ref.\cite{vWK}
\begin{equation}
\Gamma_{\alpha_1 \cdots\alpha_n}(p_1 \cdots p_n)=T_{\alpha_1 k_1}(p_1) \cdots T_{\alpha_n k_n}(p_n) \ K_{k_1 \cdots k_n}(p_1 \cdots p_n)
\label{8}
\end{equation}
where $k_i=1,2$ and $\alpha_i=R,A$. There exist different convenient choices in both bases. We follow Ref.\cite {vWK} and choose
\begin{equation}
\begin{array}{|cc|}
T_{R1}(p) & T_{R2}(p) \\ T_{A1}(p) & T_{A2}(p) 
\end{array}  =
 \begin{array}{|cc|}
N(p) & 1 \\ 1 & 0 
\end{array}
\label{9}  
\end{equation}
\begin{equation}
N(p)\ =\ {1\over 2}\coth {\beta\over 2}p^0 \ = n(p^0) + {1\over 2}
\end{equation}
Then, the relations are
\begin{equation}
\Gamma_{AA \cdots A}(p_1,p_2 \cdots p_n) = 0 = K_{11 \cdots 1}(p_1,p_2 \cdots p_n)
\label{11}
\end{equation}
\begin{equation}
\Gamma_{RA \cdots A}(p_1,p_2 \cdots p_n) = K_{21 \cdots 1}(p_1,p_2 \cdots p_n)
\label{12} \end{equation}
\begin{eqnarray}
\Gamma_{RRA \cdots A}(p_1,p_2 \cdots p_n) =  K_{221 \cdots 1}(p_1,p_2 \cdots p_n) \nonumber \\
+ \ N(p_1) \  K_{121 \cdots 1} \  +N(p_2) \  K_{211 \cdots 1}
\label{13} \end{eqnarray}
\begin{eqnarray}
\Gamma_{RRRA \cdots A}(p_1,p_2 \cdots p_n) = K_{2221 \cdots 1}(p_1,p_2 \cdots p_n) \nonumber \\
+ \ N(p_1) \  K_{1221 \cdots1} \  +  N(p_2) \  K_{2121 \cdots1} \ +  N(p_3) \  K_{2211 \cdots1} \nonumber \\
+N(p_2)N(p_3) \  K_{2111 \cdots1} \  +  N(p_1)N(p_3) \  K_{1211 \cdots1} \ + N(p_1) N(p_2) \  K_{1121 \cdots1}
\label{14} \end{eqnarray}
and so on.  Moreover, the $R/A$ basis possesses an important complex conjugate relation \cite{vW,FG1,vWK}. With the choice of Eq.(\ref{9}) and our conventions, it is
\begin{eqnarray}
{\mathcal N}( p_1 \cdots p_l) \ \Gamma(p_{1A} \cdots p_{lA}, p_{(l+1)R} \cdots p_{nR}) \nonumber \\
=(-1)^n {\mathcal N}(p_{l+1}  \cdots p_n) \ \Gamma^*(p_{1R} \cdots p_{lR}, p_{(l+1)A} \cdots p_{nA})
\label{15}
\end{eqnarray}
where
\begin{eqnarray}
{\mathcal N} (p_1 \ldots p_m)= \frac{\prod_{i=1}^m n(p_i^0)}{n(\sum_{i=1}^mp_i^0)}  \nonumber \\
 = \prod_{i=1}^m (N(p_i)+{1\over2}) \ - \   \prod_{i=1}^m (N(p_i)-{1\over2})
\label{16}
\end{eqnarray}
in particular ${\mathcal N}(p) =1$, ${\mathcal N}(p_i, p_j) = N(p_i) + N(p_j)$. For example,
\begin{eqnarray} 
\lefteqn{\Gamma_{ARR\cdots R}(p_1,p_2 \cdots p_n) =} \nonumber \\ & &
(-1)^n {\mathcal N}(p_2 \cdots p_n) \ \Gamma_{RAA \cdots A}^*(p_1, p_2 \cdots p_n)
 \label{17}\end{eqnarray}
i.e. the thermal weight are attached to the legs of type $R$, and the functions $\Gamma_{RAA\cdots A}$ are the retarded amplitudes of the response approach. \\ 

Concentrating now on the case of the ultrasoft $n$-point amplitudes, a consistent approximation is obtained if one writes in Eqs.(\ref{13}) to (\ref{17})
\begin{equation}
N(p_i)\approx{T\over p_i^0}
\end{equation}
From Eqs.(\ref{12}) and (\ref{4}), the amplitudes with one Keldysh index ``2'' are functional derivatives of the induced current $j_{\mu}(X;A)$. The amplitudes with two Keldysh indices ``2'' are functional derivatives of the function ${\mathcal K}_{\mu\nu}(X,Y;A)$ defined in Eq.(\ref{7}). $K_{22}$ is  ${\mathcal K}_{\mu\nu}(X,Y;A=0)$, one differentiation gives $K_{221}$, and so forth. An amplitude is a sum of terms made of a string of operators, and each term is linked to a specific time ordering of the gluon legs. The properties of the functional derivation are such that: i) the amplitudes are symmetric in all gluon legs of the same Keldysh index, ii) the order in $\mathbf v$ space and in color space follow the time order [from Eq.(\ref{5})], with a corresponding property in momentum space. 

The relation (\ref{13}) has an important interpretation. The retarded amplitudes $K_{211\cdots 1}$ is such that $p_1$ is always associated with the latest time. The amplitude $K_{221 \cdots1}$ is such that $X$ and $Y$ (i.e. $p_1$ and $p_2$) are latest and earliest times, or vice-versa. In relation (\ref{13}),  the terms that are present  in $K_{221 \cdots1}$  cancel the terms from  $K_{211 \cdots1} (T/p_2^0)$ such that $p_2$ is associated with the earliest time, and the terms from   $K_{121 \cdots1} (T/p_1^0)$ where $p_1$ is associated with the earliest time. The resulting terms in $\Gamma_{RRA\cdots A}$ are such that either $p_1$ or $p_2$ is associated with the latest time, while neither $p_1$ nor $p_2$ corresponds to the earliest time, a feature expected from causality. 

The constraint $\Gamma_{RR}=0$ gives the Keldysh component of the propagator
\begin{equation}
K_{22}(p_1,-p_1)={T\over p_1^0}(K_{21}-K_{12})={T\over p_1^0}[\Pi(p_{1R})-\Pi(p_{1A})]
\end{equation}
Explicit examples are 
\begin{equation}
\Pi_{\mu \nu}(p_{1R}) = m_D^2 \delta^{ab}[\int_{v.v'} v_{\mu} (v.p_1 + i \hat{C})^{-1}  {v}_{\nu}'  (p_1^0) -  g_{\mu 0} g_{\nu 0} \  ]
\end{equation}
\begin{eqnarray}
 \lefteqn{\Gamma_{RAA}=\Gamma_{\mu\nu\rho}^{abc} (p_{1R},p_{2A},p_{3A})
 = m_D^2 \int_{v,v',v"} } \nonumber    \\  & &
  [ \  i f^{acb} v_{\mu} (v.p_1+i \hat{C} )^{-1} {v}_{\rho}' (-v.p_2+i \hat{C} )^{-1}
 {v"}_{\nu}  (- p_2^0) +  ( 2 \leftrightarrow 3) \ ]
 \end{eqnarray}
 \begin{eqnarray}
 \lefteqn{K_{221}=K_{\mu\nu\rho}^{abc}(p_{1(2)},p_{2(2)},p_{3(1)})
 = m_D^2  T \int_{v,v',v"} }\nonumber  \\  & &
[ \ i f^{acb}v_{\mu}(v.p_1+i\hat{C})^{-1}{v}_{\rho}'(-v.p_2+i\hat{C})^{-1}
 {v"}_{\nu}  +  (2\leftrightarrow 1) \ ]
 \end{eqnarray}
where $(i\leftrightarrow j)$ means a term obtained by the exchange of all indices, i.e. momentum, Lorentz, colour.

The complex conjugate relation (\ref{15}) imposes multiple consistency conditions which are now examined. The complex conjugation amounts to a time reversal of each string of operators, up to a sign $(-1)^{n-1}$. The constraint
\begin{equation}
\Gamma_{RRA}=-({T\over p_1^0}+{T\over p_2^0}) \ \Gamma_{AAR}^*
\end{equation}
is satisfied very simply by the operators' strings. In $\Gamma_{RRA}$, $p_3$ is associated with the earliest time, as $p_1$ and $p_2$ cannot be, and this is also the case for the complex conjugate of $\Gamma_{AAR}$ (where $p_3$ is the latest time). In a similar way
\begin{equation}
\Gamma_{RRAA} \ ({T\over p_3^0}+{T\over p_4^0})= \Gamma_{AARR}^* \ ({T\over p_1^0}+{T\over p_2^0})
\end{equation}
In $\Gamma_{RRAA}$, $p_1$ or $p_2$ is the latest time and neither of them is the earliest time. This time ordering is equivalent to the one of $\Gamma_{AARR}^*$, where $p_3$ or $p_4$ is the earliest time and neither of them is the latest time.

Turning to relation (\ref{14}), the constraint $\Gamma_{RRR}=0$ leads to the result $K_{222}=0$, and the constraint
\begin{equation}
\Gamma_{RRRA}= \Gamma_{AAAR}^* \ T^2 ({1\over {p_1^0p_2^0}}+{1\over {p_1^0p_3^0}}+{1\over {p_2^0p_3^0}})
\end{equation}
is obeyed with $K_{2221}=0$. Then, from $\Gamma_{RRRR}=0$ one deduces that $K_{2222}=0$. All the constraints on time ordering are satisfied with the vanishing of the amplitudes with three or more Keldysh indices ``2''. One can go to the 5-point functions and check that all the constraints from Eq.(\ref{15}) are obeyed.

The set of vertices in the $R/A$ basis may be used in a perturbative expansion. For example, one-loop self-energy diagrams with loop momentum $g^2T\ln 1/g$ are made either with a pair $\Gamma_{RRA}, \Gamma_{AAR}$, or with $\Gamma_{RRAA}$ \cite{FG3} (a retarded propagator joins an $A$ leg to an $R$ leg, as it joins an outgoing $p_R$ (i.e. incoming $(-p)_A$) to an incoming  $p_R$ ). \\

One now turns to the Ward identities which are satistied by the Keldysh amplitudes. They are a consequence of the gauge covariance of the Green function (\ref{4a}) and of the conservation laws resulting from Eq.(\ref{3}), i.e.
\begin{equation}
D^{\mu} j_{\mu}^{ind} =0=D_X^{\mu} \ {\mathcal K}_{\mu\nu} = D_Y^{\nu} \ {\mathcal K}_{\mu\nu}
\end{equation}
The $n$-point retarded amplitudes are related to $(n-1)$ retarded ones \cite{FG3}. Another set of identities takes place within the subspace with two indices ``2''. For a contraction of a leg with Keldysh index ``2'', one uses relations such as
\begin{equation}
\int_v v.p_1(v.p_1+i\hat{C})^{-1}_{\ v.v'} =\int_v (v.p_1+i\hat{C}) \  (v.p_1+i\hat{C})^{-1} = 1
\end{equation}
with the help of Eq.(\ref{3}). For a contraction of a leg with index ``1'', one uses relations such as
\begin{equation}
v.p_3= (v.(p_3+p_1)+i\hat{C}) -(v.p_1+i\hat{C})
\end{equation}
Those identities allow one to cancel one propagator of each operators' string. For example, the resulting Ward identities for $K_{2211}$
are 
\begin{eqnarray}
 \lefteqn{-i p_1^{\mu} \ K_{\mu\nu\rho\sigma}^{abcd}(p_{1(2)},p_{2(2)},p_{3(1)},p_{4(1)})
=} \nonumber \\  & &
f^{acm}K_{\nu\rho\sigma}^{bmd}( p_{2(2)},(p_1+p_3)_{(2)},p_{4(1)}) +  (3\leftrightarrow 4) 
\end{eqnarray}
\begin{eqnarray}
 \lefteqn{-ip_4^{\sigma} \ K_{\mu\nu\rho\sigma}^{abcd}(p_{1(2)},p_{2(2)},p_{3(1)},p_{4(1)})
=} \nonumber \\  & &
f^{dcm}K_{\mu\nu\rho}^{abm}( p_{1(2)}, p_{2(2)},(p_3+p_4)_{(1)}) \nonumber \\ & &
 +\  [ \  f^{dam} K_{\mu\nu\rho}^{mbc}((p_1+p_4)_{(2)},p_{2(2)},p_{3(1)})  +  (1\leftrightarrow 2) \ ] 
\end{eqnarray}
One can check that these identities lead to the Ward identities which are expected in the $R/A$ basis \cite{FG3}. For example, for the $RRAA$ amplitude,  there enter the identities,  factors such as
\begin{equation}
{\mathcal N}(p_1, p_2+p_4) \approx {T\over p_1^0} +{T\over {p_2^0+p_4^0}}\  \  , \ \  {\mathcal N}(p_2, p_1+p_3) \approx {T\over p_2^0} +{T\over {p_1^0+p_3^0}}  \ \ .
\end{equation} \\

To conclude,  we summarize the argument. The induced current $j_{\mu}$ is the generating function of the retarded ultrasoft amplitudes, and, in particular, of the full set of two- and three-point functions. With the approximation   $N(p_0)\approx  T/p_0$ (the usual companion to the soft amplitudes), one deduces, in the Keldysh basis, the expressions for $K_{22}$ , $K_{221}$ and the relation $K_{222}=0$ . Then, from $K_{22}$ , $K_{221}$ , gauge symmetry suggests to step across to the generating function $K_{\mu \nu}$ of all amplitudes with two Keldysh indices 2, i.e. the Green function (\ref {4a}) enters $K_{\mu \nu}$. Moreover, general constraints exist on the set of $n$-point amplitudes. They are stated in the $R/A$ basis, and they allow to advance towards the full set of $n$-point amplitudes. One obtains the result that all amplitudes with more than two indices 2 vanish.

This set of ultrasoft amplitudes may be derived from an effective action in the Keldysh basis. In addition to the field $A_{\mu}(X)$, of type 1, one introduces a field $B_{\mu}(X)$ of type 2. The effective action's potential term is
\begin{equation}
\Gamma^V(A,B)=Tr[\ \int_xJ_{\mu}(X ; A)\ B^{\mu}(X) \ + \int_{x,y}{1\over2} B^{\mu}(X) \ K_{\mu \nu}(X,Y ; A) \ B^{\nu}(Y) \ ]
\end{equation}
with $\int_x=\int_{-\infty}^{+\infty} dt \int d^3 x$ . The role of the field $B$ is clarified if one recalls the following properties of the Keldysh basis \cite{Chou} : (i) the Legendre transform reads
$ E(J_1, J_2) = - \int (J_1.B + J_2.A) - \Gamma(A,B)  $ (ii) the source $J_2$ generates the observables in $A$, (iii) $B=0$ amounts to $J_2=0$,  (iv) the relation $B=0$ says that the field A takes the same value on the forward and backward branches of the time contour in the Closed-Time-Path basis.


\begin{thebibliography}{40}

\bibitem{Bod} D. B\"{o}deker, Phys. Lett. B 426 (1998) 351;  Nucl. Phys. B 559 (1999) 502;  B 566 (2000) 402.

\bibitem{BI1} J. P. Blaizot and E. Iancu, Nucl. Phys. B 570 (2000) 326; B 557 (1999) 183.
 
\bibitem{AY} P. Arnold, D. Son and L.G. Yaffe, Phys. Rev. D 59 (1999) 105020; D 60 (1999) 025007.

\bibitem{BI3} J. P. Blaizot and E. Iancu, Nucl. Phys. B 417 (1994) 608; B 434 (1995) 662; hep-ph/0101103.

\bibitem{FG3} F. Guerin, Phys. Rev. D 63 (2001) 045017.

\bibitem{MLB} M. Le Bellac, {\it Thermal Field Theory} (Cambridge University Press, Cambridge, 1996).

\bibitem{Au1} P. Aurenche and T. Becherrawy, Nucl. Phys. B 379 (1992) 259. 

\bibitem{Chou} K.Chou, Z. Su, B. Hao, and L. Yu, Phys. Rep. 118 (1985) 1.

\bibitem{vWK} M. van Eijck, R. Kobes, and Ch. G. van Weert, Phys. Rev. D 50 (1994) 4097. 

\bibitem{vW} C. Van Eijck and Ch. G.  Van Weert, Phys. Lett. B 278 (1992) 305. 

\bibitem{FG1} F. Guerin, Nucl. Phys. B 432 (1994) 281. 


\end{thebibliography}
\end{document}